\documentclass[twocolumn, secnumarabic, amssymb, amsmath,nobibnotes, aps, prl,nofootinbib]{revtex4}
\begin{document}
\title{Entropies of Mixing and the Lorenz Order}
\author{B. H. Lavenda}
\email{bernard.lavenda@unicam.it}
\affiliation{Universit$\grave{a}$ degli Studi, Camerino 62032 (MC) Italy}
\date{\today}
\newcommand{\sumn}{\sum_{i=1}^{n}\,}
\newcommand{\sumk}{\sum_{i=1}^{k}\,}
\newcommand{\half}{\mbox{\small{$\frac{1}{2}$}}}
\newcommand{\third}{\mbox{\small{$\frac{1}{3}$}}}\newcommand{\twothirds}{\mbox{\small{$\frac{2}{3}$}}}
\begin{abstract}
Entropies of mixing can be derived directly from the parent distributions of extreme
value theory. They correspond to pseudo-additive entropies in the case of Pareto and power 
function distributions, while to the Shannon entropy in the case of the exponential distribution.
The former tend to the latter when their shape parameters tend to infinity and zero, 
respectively. Hence processes whose entropies of mixing are pseudo-additive entropies majorize,
in the Lorenz order sense, those whose entropy is the Shannon entropy. In the case of the
arcsine distribution, maximal properties of regular polygons correspond to maximum entropy of
mixing.
\end{abstract}
 
\maketitle
\section{Majorization and the Lorenz order}
The employment of entropies as measures of inequality dates back a long way \cite{MO}. 
Majorization, as well as Lorenz ordering, consists of a partial order of $n$-tuples 
that lie on the postive orthant of $n$-dimensional Euclidean space, $\mathbb{R}_{+}^n$ \cite{A}. Comparison
of different populations can be made with respect to their diversity. Arranging two sets of $n$-tuples,
$x_1,x_2,\ldots,x_n$ and $y_1,y_2,\ldots,y_n$ in increasing order, $x_{1:n}<x_{2:n}<
\ldots<x_{n:n}$ and $y_{1:n}<y_{2:n}<\ldots<y_{n:n}$, a vector $\vec{y}$ is said to
majorize another vector $\vec{x}$, $\vec{y}\succ\vec{x}$, if $\sumk y_{i:n}<\sumk x_{i:n}$ 
for $k=1,2,\ldots,n-1$, and $\sumn y_{i:n}=\sumn x_{i:n}$. Weak majorization converts the 
last inequality into an inequality like the former one \cite[pp. 9ff]{MO}. Expressed in
words, this says that $\vec{x}$ does not exhibit more inequality than $\vec{y}$.\par
The characteristic Lorenz bow-shaped curves, indicating that there are many more poor people
than rich ones, are obtained by plotting the points 
$\left(k/n,\sumk x_{i:n}/\sumn x_{i:n}\right)$ and
$\left(k/n,\sumk y_{i:n}/\sumn y_{i:n}\right)$, and using linear interpolation. If the 
Lorenz curve of $\vec{x}$ is obtained by raising the $\vec{y}$ curve at one or several
points, then $\vec{y}\succ\vec{x}$. Lorenz curves are convex, and prosaically speaking 
\lq\lq as the bow is bent so inequality (or the concentration of wealth) increases\rq\rq.
\par
Majorization involves a partial ordering of vectors in $\mathbb{R}_{+}^n$. If $n$ is so 
large as to make partial ordering impractical, sufficient conditions for  majorization can
be obtained by indentifying a Schur-convex function. Oddly enough, Schur's theory of
convex functions preceded the theory of majorization by about a decade. A function $f$ is
said to be Schur-convex when $\vec{y}\succ\vec{x}\Rightarrow f(y)\ge f(x)$. In particular
contexts, like thermodynamics and ecology, measures of inequality should be maximum when
all the components are equal; here enters the notion of entropy. In so doing, we transfer
our attention from a Schur-convex to a Schur-concave function. Until now there has been
no unique relation between the entropy employed and the underlying probability distribution.
The maximum entropy formalism chooses the form of the entropy and derives an exponential
family of distributions by invoking particular constraints through a variational procedure. Here, we shall show that the
probability distribution determines the form of the entropy uniquely via the Lorenz 
function.\par
Lorenz order has meaning for continuous, as well as discrete, probability distributions, 
but only those that have support on non-negative reals and have finite means. In fact,
the Lorenz function is defined in terms of the normalized incomplete mean [cf.  (\ref{eq:Lorenz}) below]. For certain distributions, like the Pareto one, the condition that
the mean be finite imposes certain restrictions on the shape parameter that will appear in
the characteristic exponents in the corresponding entropies of mixing. Although it has
been lamented that the list of Lorenz functions that can be obtained in closed form is
remarkably short \cite[p. 34]{A}, we will show that they contain all the parent 
distributions of extreme value theory as well as the arcsine law. Since there are only 
three families of exteme value distributions, this will place an upper limit to the number
of Lorenz curves that can be obtained in closed form, apart from those obtained through
geometrical inequalities, like the arcsine distribution.\par In fact, the appearance of extreme value theory was not unexpected since it enters naturally into order statistics. Although
ordering destroys the independence and common distribution of the initial random variables, it does lead to some very simple results for extreme values when the sample size is allowed to increase without limit. That is, if we consider the $k$th order statistic in a sample
of size $n$ and let both $k$ and $n$ tend to infinity such that $k/n=\mbox{const}.$, the
asymptotic distributions of the quantiles are obtained \cite{SO}. Rather, if $k$ is fixed
and the population size is allowed to increase without limit, the asymptotic distributions
of extreme values result. The parent distributions which lie in the domain of attraction
of the extreme value distributions for largest value are the exponential, Pareto, and power
tail distributions. These distributions are attracted to the double exponential, or what is commonly referred to as the Gumbel,
distribution, the Fr\'echet distribution, and the reversed Weibull distribution, 
respectively. We will, instead, deal with the power function distribution which generates
the Weibull distribution for smallest value since it has the same entropy of mixing and is 
simpler to handle. \par
Although these families of extreme value distributions appear as distinct, we will show that
their entropies of mixing, which are derived from the parent distributions, all tend to a single entropy of mixing as the shape parameter approaches a certain limit. On the same Lorenz plot, we shall see that the entropies of mixing are cigar-shaped curves, and the cigar-shaped curve of the entropy of mixing of the exponential distribution is nested in
those of the Pareto and power function entropies of mixing. That is to say that the entropy
of mixing of the exponential distribution will be smaller than those of the Pareto and power
function distributions. The entropies of mixing can thus be used in the process of Lorenz
ordering and are Schur-concave functions that are determined uniquely by their Lorenz 
functions.\par
The probability distributions  determine the Lorenz function up to a scale 
transformation \cite[p. 32]{A}, and the difference between the tail of the Lorenz function
and the Lorenz function itself determines the entropy of mixing. This is analogous to the
difference between the upper and lower quartiles which is used as a measure of dispersion
of the distribution \cite{C}. The difference in the Lorenz functions is a measure of 
uncertainty, and uncertainty will be greatest when the probabilities are equal. This
results in the maximum entropy of mixing.\par As a measure of inequality, the entropies of
mixing are comparable to the Gini index, which is defined as twice the area between
the Lorenz curve and the $45^0$ line, or the unbent bow, representing an equal distribution
of income for all individuals in the population. Whereas the Gini index is a global
criterion of inequality, the entropies of mixing are local.\par
The entropies of mixing obtained from the parent distributions are not unfamiliar in
the information sciences. Those that are obtained from the Pareto and power function 
distributions will be recognized as pseudo-additive entropies (pae) \cite{AD}, which have
been popularized by Tsallis in the physical sciences \cite{T}. In well-defined limits of
the shape parameter of the distributions, these pae transform into the Shannon entropy
corresponding to the exponential distribution.\par
Another expression for the Lorenz function that can be obtained in closed form corresponds
to the arcsine law. The maximum entropy of mixing of this distribution will be shown to
coincide with the maximal properties of regular polygons. In this way the maximum uncertainty
expressed in terms of entropy of mixing is related to the geometrical conditions for greatest
area and perimeter of regular polygons.\par
\section{Lorenz Function and Entropy of Mixing}
The Lorenz curve plots the percentage of total income by various fractions of the population 
ordered in increasing size of their incomes. Although Lorenz functions have been known in 
parametric form for quite some time, it is only relatively recently that a definition in terms of
a single equation, rather than in terms of two equations, has been advanced \cite{G}.\par
The conventional definition of the Lorenz curve is the following \cite[Sec. 2.25]{SO}. Given 
the cumulative distribution on $\mathbb{R}_{+}^n$ as
\begin{equation}
p=F(x)=\int_0^x\,f(t)\,dt \label{eq:F}
\end{equation}
one solves for $x$ and writes the Lorenz function as the normalized, incomplete mean
\[
L(p)=\frac{1}{\mu}\int_0^x\,t f(t)\,dt,
\]
where $\mu=\int_0^{\infty}t\,dF(t)$ is the finite mean, and $x$ and $p$ are related by (\ref{eq:F}).\par An alternative formulation \cite{G} uses the
fact that the solution to (\ref{eq:F}) is
\[
F^{-1}(t)=\sup_{x}\left\{x:F(x)\le t\right\}, \;\;\;\;\;\;\;\; 0<t<1.
\]
In terms of the inverse distribution function, the mean is given by the Riemann integral
\begin{equation}
\mu=\int_0^1\,F^{-1}(t)\,dt. \label{eq:mu}
\end{equation}
And in terms of $F^{-1}$, the Lorenz function is defined as \cite{G}
\begin{equation}
L(p)=\frac{1}{\mu}\int_0^p\,F^{-1}(t)\,dt. \label{eq:Lorenz}
\end{equation}
\par
The Lorenz function, (\ref{eq:Lorenz}), is non-decreasing and convex on $[0,1]$, with $L(0)=0$
and $L(1)=1$. Most importantly, the Lorenz function determines the distribution function, $F$, up
to a scale transformation
\begin{equation}
L^{\prime}(p)=F^{-1}(p)/\mu, \label{eq:key}
\end{equation}
since $F^{-1}$ determines $F$. Moreover, if $L$ is twice 
differentiable then $F$ will have a finite positive density given by
\[f(x)=\left[\mu L^{\prime\prime}\left(F(x)\right)\right]^{-1}.\]
\par
In fact, von Mises's (sufficient) conditions for a  distribution to be in the domain of attraction of an extreme value distribution can be very succinctly expressed in terms of the ratios of the derivatives of the Lorenz function, without having to take the limit where the variate tends to its right end point, $\sup_{x}\left\{x:F(x)<1\right\}$. For distributions unlimited on the right, like the Pareto law, $F^{-1}(1)=\infty$, and
\[\alpha(1-p)=L^{\prime}(p)/L^{\prime\prime}(p).\]
For those that are limited on the right, $F^{-1}(1)<\infty$, and
\[\beta(1-p)=\frac{L^{\prime}(1)-L^{\prime}(p)}{L^{\prime\prime}(p)}.\]
Finally, for the exponential distribution the condition is
\[1-p=1/L^{\prime\prime}(p).\]
\par
The tail of the Lorenz function can likewise be defined as
\begin{equation}
\bar{L}(p):=1-L(1-p)=\frac{1}{\mu}\int_{1-p}^1\,F^{-1}(t)\,dt. \label{eq:tail}
\end{equation}
Since $L(p)$ is convex, $\bar{L}(p)$ is concave. The difference between (\ref{eq:tail}) and
(\ref{eq:Lorenz}) gives the  entropy of mixing as
\begin{equation}
S(p)=\frac{\bar{L}(p)-L(p)}{1-2^{-D}}=\frac{1-L(1-p)-L(p)}{1-2^{-D}}, \label{eq:S}
\end{equation}
after having been suitably normalized, where the exponent $D$ in the normalizing denominator is given in terms of the shape parameter of the distribution, and it can be
both positive or negative.
\section{Entropies of mixing of parent distributions}
The classic example of a Lorenz function is obtained from the untranslated Pareto distribution
\begin{equation}
F(x)=1-\left(\frac{x_0}{x}\right)^{\alpha}, \label{eq:Pareto}
\end{equation}
describing the distribution of salaries above $x_0$. In order for there to be a finite mean, 
$\alpha>1$. Let $x=F^{-1}(p)$ so that
\[
F^{-1}(p)=x_0(1-p)^{-1/\alpha}.\]
Since the derivative of the Lorenz function is related to the inverse distribution function according
to (\ref{eq:key}) we obtain
\begin{equation}
L(p)=1-(1-p)^{(\alpha-1)/\alpha}, \label{eq:L-Pareto}
\end{equation}
upon integrating,  evaluating the limits as given in (\ref{eq:Lorenz}), and using the expression $\mu=\frac{\alpha}{\alpha-1}x_0$ for the mean.\par The tail of the Lorenz
curve, defined in (\ref{eq:tail}), will be given by the expression
\begin{equation}
\bar{L}(p)=p^{(\alpha-1)/\alpha} \label{eq:barL-Pareto}. 
\end{equation}
Whereas (\ref{eq:L-Pareto}) represents the
fraction of total income possessed by the lowest $p$-th fraction of the population, (\ref{eq:barL-Pareto}) represents the fraction of total income held by the highest $p$-th fraction.
\par
The difference between the Lorenz functions, (\ref{eq:barL-Pareto}) and (\ref{eq:L-Pareto}), yield an entropy of mixing
\begin{equation}
S_{-}(p)=\frac{p^{(\alpha-1)/\alpha}+(1-p)^{(\alpha-1)/\alpha}-1}{2^{1/\alpha}-1}. 
\label{eq:S-Pareto}
\end{equation}
This is precisely a normalized pae of information theory \cite{AD}. The entropy of mixing (\ref{eq:S-Pareto}) reflects the symmetry between the top and bottom order statistics. The difference in Lorenz functions is a measure of uncertainty, just like the difference in quartiles is a measure of dispersion. The maximum uncertainty occurs when $p=\half$ for which $S_{-}(\half)=1$.\par
As a second example, consider the exponential distribution
\begin{equation}
F(x)=1-e^{-x/\mu}. \label{eq:exp}
\end{equation}
The Lorenz functions are $L(p)=(1-p)\log (1-p)+p$ and $\bar{L}(p)=p-p\log p$ so that the corresponding normalized entropy of mixing is 
\begin{equation}
S_0(p)=\frac{-p\log p-(1-p)\log(1-p)}{\log 2}, \label{eq:Shannon}
\end{equation}
which will be appreciated as the (normalized) Shannon entropy of information theory. The normalization has
been chosen such that the entropy is maximal for $p=\half$, viz., $S(\half)=1$ \cite[p. 34]{AD}.\par
The entropy of mixing of the Pareto law, (\ref{eq:S-Pareto}), transforms into the entropy of
mixing of the exponential distribution, (\ref{eq:Shannon}), in the limit as $\alpha\uparrow\infty$. 
Since the exponential law, (\ref{eq:exp}), is the parent distribution of the double 
exponential distribution of largest value, this distribution can be seen to play a limiting role
in regard to the two other extreme value distributions, the Fr\'echet and reversed Weibull 
distributions.\par
In fact, the entropy of mixing associated with the power function distribution,
\begin{equation}
F(x)=\left(\frac{x}{x_0}\right)^{\beta}, \label{eq:power}
\end{equation}
which is limited on the right by $x_0=\sup_{x}\left\{x:F(x)<1\right\}$, will also tend to the Shannon entropy, 
(\ref{eq:Shannon}), in the limit where the positive shape parameter $\beta\downarrow0$. The
Lorenz functions are $L(p)=p^{(\beta+1)/\beta}$ and $\bar{L}(p)=1-(1-p)^{(\beta+1)/\beta}$ so
that the entropy of mixing is \footnote{The same entropy of mixing would have been obtained had
we treated the tail of the power function,
\[F(x)=1-\left(1-x\right)^{\beta},\] 
which generates the reverse Weibull distribution for largest value. This is the distribution of
lengths obtained when $\beta$ independent and random chosen points partition the interval,
$\overline{0,1}$ into $\beta+1$ intervals \cite{F}. The entropy of mixing remains the same
even though $L(p)$ and $\bar{L}(p)$ have more complicated forms, viz., $\bar{L}(p)=(\beta+1)p-\beta p^{(\beta+1)/\beta}$ and $L(p)=(\beta+1)p+\beta[(1-p)^{(\beta+1)/\beta}-1]$, where the mean, $\mu=(\beta+1)^{-1}$, has been used to evaluate the expressions. Both vanish at $p=0$, and are unity at $p=1$. Their difference  is precisely (\ref{eq:S-power}) when properly normalized; the normalizing denominator being $\beta\left(1-2^{-1/\beta}\right)$.}
\begin{equation}
S_{+}(p)=\frac{1-(1-p)^{(\beta+1)/\beta}-p^{(\beta+1)/\beta}}{1-2^{-1/\beta}}, \label{eq:S-power}
\end{equation}
where we have set $D=1/\beta$ in (\ref{eq:S}). In the limit $\beta\downarrow0$, the pae 
(\ref{eq:S-power}) transforms into the Shannon entropy, (\ref{eq:Shannon}).\par
Therefore, as the shape parameter of the Pareto distribution (\ref{eq:Pareto}) increases, or
the shape parameter of the power function distribution (\ref{eq:power}) decreases, there is a
decrease in inequality as measured by Lorenz ordering. The criterion of majorization, or Lorenz
ordering, is given by the inequality of the entropies of mixing
\begin{equation}
S_{\pm}(p)>S_0(p). \label{eq:Sineq}
\end{equation}
The inequality says that the Lorenz curves of the exponential distribution are nested in the 
Lorenz curves of the Pareto and power function distributions. In other words, any process whose
distribution is exponential does not exhibit any more inequality in the Lorenz sense than a
process governed by either the Pareto or power function distributions.\par
The entropies of mixing, (\ref{eq:S-Pareto}) and (\ref{eq:S-power}), are separable, Schur-
concave functions. Considering (\ref{eq:S-Pareto}), it can be written as the sum
\[
S_{-}(p)=\sum_{i=1}^2\sigma(p_i),
\]
where
\[
\sigma(p_i)=\frac{p_i^{(\alpha-1)/\alpha}-p_i}{2^{1/\alpha}-1}. \]
Since $\sigma^{\prime\prime}(p_i)<0$, we can apply Jensen's inequality in the form
\[
\sigma\left(\sum_{i=1}^2 q_i p_i\right)>\sum_{i=1}^2 q_i\sigma(p_i),
\]
or, equivalently,
\begin{equation}
\sum_{i=1}^2 q_i p_i>\left(\sum_{i=1}^2 q_i p_i^{(\alpha-1)/\alpha}\right)^{\alpha/(\alpha-1)}. 
\label{eq:means}
\end{equation}
The inequality in (\ref{eq:means}) follows from the fact that power means are increasing functions
of their order \cite{HLP} since $\alpha>1$. In the limit as $\alpha\rightarrow 1$, (\ref{eq:means}) becomes the 
arithmetic-geometric mean inequality. \par
In particular, if we set $q_i=\half$, the entropy of mixing of the Pareto distribution is
seen to be bounded from above, viz.,
\[S_{-}(p)=\sum_{i=1}^2\,\sigma(p_i)\le 2\sigma\left(\half\sum_{i=1}^2\, p_i\right)=2\sigma\left(\half\right)=1.
\]
This shows that $S_{-}\left(\half\right)=1$ is indeed maximal.\par
From the above discussion if follows that if $y\ _{L}\!\!\succ x$, i.e. that $x$ does not show more
inequality than $y$, the entropy of mixing of $y$ is greater than that of $x$. If $x$ and $y$ 
have two Pareto distributions with shape parameters $\alpha_x>\alpha_y$, then
$S_{-}^{(y)}(p)>S_{-}^{(x)}(p)$. The entropy of mixing decreases as the shape parameter of
the Pareto distribution increases, corresponding to a decrease in inequality as measured by the Lorenz ordering.
\section{Maximal properties of regular polygons and maximum entropy of mixing}
A closed form expression for the Lorenz functions also exists for the arcsine distribution, which lies outside the traditional formulation of extreme value distributions. The arcsine law
\begin{equation}
F(x)=\sqrt{\frac{2}{\pi}}\sin^{-1}\sqrt{x} \label{eq:arc}
\end{equation}
is a symmetric distribution where the probabilities at the extremes, $x=0$, and $x=1$, are the greatest. The central term has the smallest probability even though it coincides with the mean value, $\mu=\half$. This goes against intuition which equates mean and most probable values.\par
The Lorenz functions
\[L(p)=p-\frac{\sin\pi p}{\pi},\]
and
\[\bar{L}(p)=p+\frac{\sin(1-p)\pi}{\pi},\]
reflect the symmetry of the arcsine distribution. Their difference gives the
entropy of mixing as
\begin{equation}
S(p)=\half\left[\sin\pi p+\sin(1-p)\pi\right]=\sin\pi p, \label{eq:S-arc}
\end{equation}
upon normalization, which is concave and maximal.\par The entropy of mixing, (\ref{eq:S-arc}), is easily generalized to a set of $n$ probabilities, viz.,
\[
S(p)=\frac{1}{n}\sumn \sin\pi p_i, \]
where the set $\{p_i\}$ is assumed to form a complete distribution. Since $(\sin\vartheta)^{\prime\prime}<0$ and $0<\vartheta<\pi$, Jensen's inequality gives \cite[p. 99]{HLP}
\begin{equation}
\sin\left(\pi\sumn p_i\big/n\right)>\frac{1}{n}\sumn\sin\pi p_i, \label{eq:Jensen}
\end{equation}
unless all the $p_i$ are equal. Now since the $p_i$ are assumed to form a complete distribution, (\ref{eq:Jensen}) reduces to
\begin{equation}
n\sin\left(\frac{\pi}{n}\right)>\sumn\sin\pi p_i. \label{eq:Jensen-bis}
\end{equation}
\par
The left side of (\ref{eq:Jensen-bis}) is half the perimeter of a regular polygon of $n$ sides inscribed in a circle of unit radius. Let $O$ be the center of the circle and $P_0,P_1,\ldots,P_n$  the vertices of a polygon that lie on the circle, where $P_0=P_n$ fixed but $P_1,P_2,\ldots,P_{n-1}$ can vary.  If the angle $P_{i-i}OP_i$ is identified with $\pi p_i$, then (\ref{eq:Jensen-bis}) asserts that both the perimeter and area of the polygon are greatest when the polygon is regular, viz., $P_{i-1}P_i=n$ for $i=1,2\ldots,n-1$. Hence, the familiar maximal properties of regular polygons coincide with the maximum entropy of mixing which occurs when all the probabilities are equal, $P_{i-1}OP_i=\pi/n$ for $i=1,2,\ldots,n-1$.

\end{document}